\documentclass[a4paper]{jpconf}

\usepackage{xspace}
\usepackage{psfrag}
\usepackage{epsfig}
\usepackage{graphicx}
\usepackage{amssymb}

\newcommand{\CnuB}{C\(\nu\)B\xspace}
\newcommand{\UHEnu}{UHE\(\nu\)\xspace}
\newcommand{\slache}[1]
{#1\!\!\!\!/}

\begin{document}
\title{Thermal effects on the absorption of ultra-high energy neutrinos by the cosmic neutrino 
background}

\author{J C D'Olivo, L Nellen, S Sahu and V Van Elewyck}

\address{Instituto de Ciencias Nucleares, Universidad Nacional Aut\'onoma de M\'exico}

\ead{vero@nucleares.unam.mx}

\begin{abstract}
We use the formalism of finite-temperature field theory to study the interactions of ultra-high 
energy (UHE) cosmic neutrinos with the background of relic neutrinos and to 
derive general expressions for the UHE neutrino transmission probability. This approach allows us 
to take into account the thermal effects introduced by the momentum distribution of the relic 
neutrinos. We compare our results with the approximate expressions existing in the literature and 
discuss the influence of thermal effects on the absorption dips in the context of favoured 
neutrino mass schemes, as well as in the case of clustered relic neutrinos.
\end{abstract}

\section{Introduction}

The interaction of cosmic neutrinos at ultra-high energies (\UHEnu) with the cosmological 
background of relic (anti) neutrinos (\CnuB) has been proposed as an indirect way of observing the 
\CnuB \cite{UHEnuCnuB}. 
Provided adequate sensitivity and energy resolution of the detectors, the observation in the 
\UHEnu flux of absorption lines associated with the resonant production of a Z boson 
($\nu\bar{\nu} \rightarrow Z \rightarrow f \bar{f}$) could allow the determination of the absolute 
neutrino masses \cite{mass}. The shape and depth of these absorption dips may also reflect 
features of the distribution of \UHEnu sources and their emission spectrum \cite{ringwald,quigg}. 
Most of the work in the literature describe the \UHEnu -\CnuB interactions assuming that relic 
neutrinos are at rest. The effects of thermal motion in the \CnuB (whose present temperature is 
$\approx 1.69 \times 10^{-4}$ eV), however, become relevant as soon as the momentum of the relic 
neutrinos gets comparable to their mass. In this work, we take into account the thermal effects on  
the dominant (resonant) contribution to the neutrino damping by using finite-temperature field 
theory (FTFT) \cite{ourpaper}. In section \ref{sec:trans} we present the \UHEnu transmission 
probability across the \CnuB for various mass schemes, and in section \ref{sec:clustering} we 
briefly discuss the impact of thermal motion on the absorption in relic neutrino clusters.

\section{Transmission probability of \UHEnu across the \CnuB}
\label{sec:trans}
For an \UHEnu with four-momentum $k^\mu = (\mathcal{E}_{_K},\vec{K})$ and mass $m_\nu$ travelling 
across the \CnuB, the equation of motion reads $(\slache{k} - m_\nu - \Sigma)\  \psi = 0$. The 
self-energy $\Sigma$ embodies the effects of the surrounding neutrino medium, coming in this case 
from the exchange of a Z boson. The corresponding dispersion relation is given by $\mathcal{E}_K = 
\mathcal{E} _r - i\frac{\gamma}{2}$, where $\mathcal{E}_r$ and $\gamma$ are functions of $K$. In 
the real-time formalism of FTFT, one calculates $\Sigma$ from the corresponding Feynman diagram,  
having in mind that the vertices are doubled respect to the standard theory, and that the 
propagators of the Z boson (near the resonance) and of the relic neutrino become $2 \times 2$ 
matrices. The relic neutrino propagator also depend on the functions $f_\nu(P)$ and 
$f_{\bar{\nu}}(P)$ characterizing the momentum distribution of (anti) neutrinos in the thermal 
bath. These functions take the simple relativistic Fermi-Dirac form $f_\nu(P) = 
f_{\bar{\nu}}(P)=1/(e^{P/T_\nu}+1)$, where
$T_{\nu}$ is the temperature of the \CnuB and we have neglected the chemical potential. 
 
The damping factor $\gamma$, which governs the attenuation of the \UHEnu flux across the 
background of relic neutrinos, is directly related to the imaginary part of the self-energy 
$\Sigma_i$ \cite{dolivo}. In the approximation that the \UHEnu are ultrarelativistic and that we 
can neglect the background effects on their energy ($\mathcal{E} _r \simeq K$), we may write the 
damping as (see \cite{ourpaper} for the detailed calculation)
\begin{equation}
\label{eq:gammaUR} \gamma_{\nu\bar{\nu}}(K) = 
-\frac{1}{K}\left.\mathop{\mathrm{Tr}}(\slache{k}\Sigma_i)\right|_{\mathcal{E}_r=K}
= \int_0^\infty \frac{dP}{2\pi^2}
\ P^2 \ f_{\bar{\nu}} (P) \ \sigma_{\nu\bar{\nu}} (P,K).
\label{eq:gamma}
\end{equation}
For $m_\nu \ll M_Z,K$ and neglecting terms of order $\Gamma_Z^2/M_Z^2$, we have 
\begin{eqnarray}
\sigma_{\nu\bar{\nu}}(P,K) &=& \frac{2\sqrt{2}G_\mathrm{F}\Gamma_Z M_Z}{2KE_p}
\left\{ 1 + \frac{M_Z^2}{4KP} \ln\left(\frac{4K^2(E_p + P)^2 -
4M_Z^2K(E_p+P)+M_Z^4}{4K^2(E_p - P)^2 -
4M_Z^2K(E_p-P)+M_Z^4}\right)\right. \nonumber \\
&& \nonumber\\
&& \left. + \frac{M_Z^3}{4KP \Gamma_Z}\left[\arctan\left(\frac{2K(E_p+P)-M_Z^2}{\Gamma
M_Z}\right) -\arctan\left(\frac{2K(E_p-P)-M_Z^2}{\Gamma
M_Z}\right)\right]\right\}. \label{eq:sigma}
\end{eqnarray}
where $E_p = \sqrt{P^2 + m_\nu^2}$ is the energy of the relic neutrino. Taking the limit of 
eq.(\ref{eq:sigma}) for $P \rightarrow 0$, one recovers the approximated cross-section which is 
used for relic neutrinos at rest, with the Z peak at the "bare" resonance energy $K_{res} = 
M_Z^2/(2m_\nu)$.  

 The transmission probability for an \UHEnu emitted at a redshift $z_s$ to be detected on Earth 
with an energy $K_0$ is obtained by integrating the damping along the \UHEnu path, taking into 
account that both the \UHEnu energy and in the \CnuB temperature are redshifted:
\begin{equation}
\label{eq:PTredshift}
P_\mathrm{T}(K_0,z_\mathrm{s})=\exp\left[{-\int_0^{z_\mathrm{s}} \frac{dz}{H(z)(1+z)} 
\gamma_{\nu\bar{\nu}}(K_0(1+z))}\right],
\end{equation} 
where $H = H_0\ \sqrt{0.3(1+z)^3 +0.7}$ is the Hubble factor. We determined the absorption dips in 
$P_\mathrm{T}$ for $m_\nu$ ranging from $10^{-1}$ eV to $10^{-4}$ eV (see \cite{ourpaper}) and 
found that the approximation of relic neutrinos at rest breaks down as soon as $m_\nu / T_\nu 
\lesssim 10^{-2}$: the absorption dips get broadened and shift to lower \UHEnu energies, and the 
effect increases with $z_s$. Fig.~1 shows the average transmission probability $\bar{P}_T$ 
obtained by summing on the neutrino flavours \cite{ringwald}, for mass patterns compatibles with 
the currently favoured 3-neutrinos mass schemes \cite{bell}. The absorption dip corresponding to 
the smallest mass is almost always washed out and only contributes to further broadening the 
absorption dip at high energies. At very large $z_s$, the merging of the two other absorption dips 
in the normal hierarchy scheme makes it more difficult to differentiate from the inverted 
hierarchy one. We conclude that thermal effects significantly affect the \UHEnu transmission 
probability well before the relic neutrinos become relativistic. This complicates the extraction 
of $m_\nu$ and $z_s$ from the startpoint and endpoint of the absorption dip (which, in the 
approximation $P=0$, were located at $K_0=K_{res}/(1+z_s)$ and $K_0=K_{res}=M_Z^2/(2m_\nu)$ 
respectively).

\begin{figure}
\label{fig:transmission}
 \psfrag{27}[c]{\tiny \phantom{n} \raisebox{0cm}{$10^{27}$}}
 \psfrag{26}[c]{\tiny \phantom{n} \raisebox{0cm}{$10^{26}$}}
 \psfrag{25}[c]{\tiny \phantom{n} \raisebox{0cm}{$10^{25}$}}
 \psfrag{24}[c]{\tiny \phantom{n} \raisebox{0cm}{$10^{24}$}}
 \psfrag{23}[c]{\tiny \phantom{n} \raisebox{0cm}{$10^{23}$}}
 \psfrag{22}[c]{\tiny \phantom{n} \raisebox{0cm}{$10^{22}$}}
 \psfrag{21}[c]{\tiny \phantom{n} \raisebox{0cm}{$10^{21}$}}
 \psfrag{20}[c]{\tiny \phantom{n} \raisebox{0cm}{$10^{20}$}}
 \psfrag{1.}[c]{\tiny \phantom{} \raisebox{0.1cm}{$1$}}
 \psfrag{0.8}[c]{\tiny \phantom{} \raisebox{0.1cm}{$0.8$}}
 \psfrag{0.6}[c]{\tiny \phantom{} \raisebox{0.1cm}{$0.6$}}
 \psfrag{0.4}[c]{\tiny \phantom{} \raisebox{0.1cm}{$0.4$}}
 \psfrag{0.2}[c]{\tiny \phantom{} \raisebox{0.1cm}{$0.2$}}
 \psfrag{0.}[c]{\tiny \phantom{} \raisebox{0.1cm}{$0$}}
 \psfrag{X}[c]{\tiny \raisebox{-0.5cm}{$K_0 [\mathrm{eV}]$}}
 \psfrag{Y}[c]{\tiny {$_{\bar{P}_\mathrm{T}}$}}

 \psfig{file=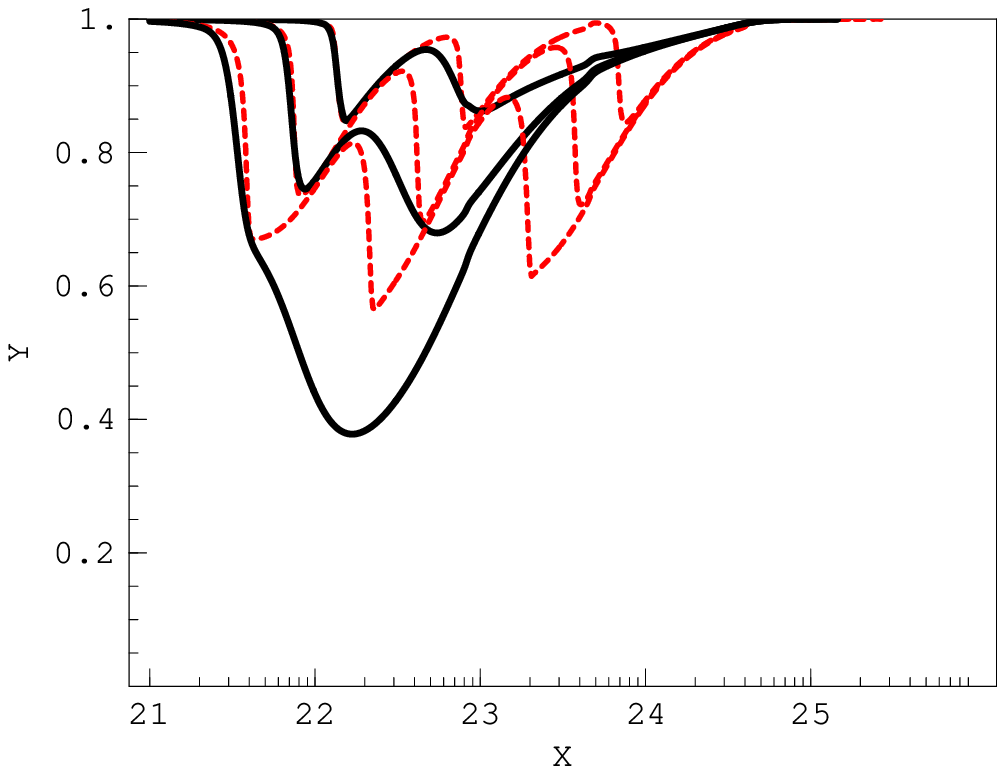,width=0.22\textwidth,angle=0}
 \hfill
 \psfig{file=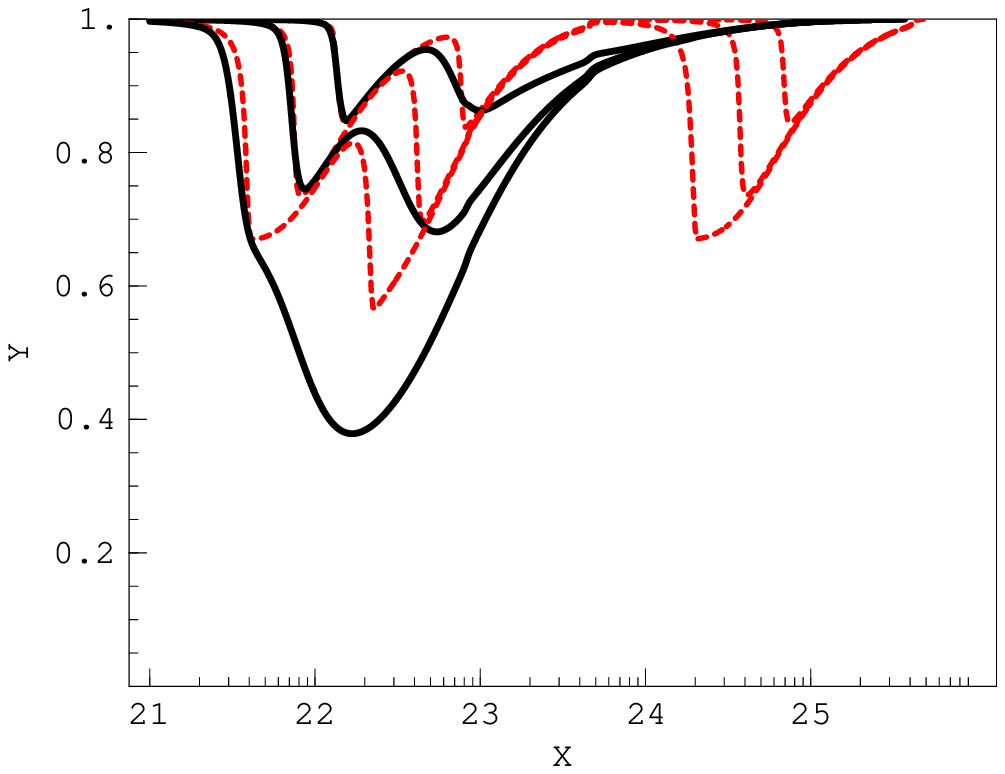,width=0.22\textwidth,angle=0} 
 \hfill
 \psfig{file=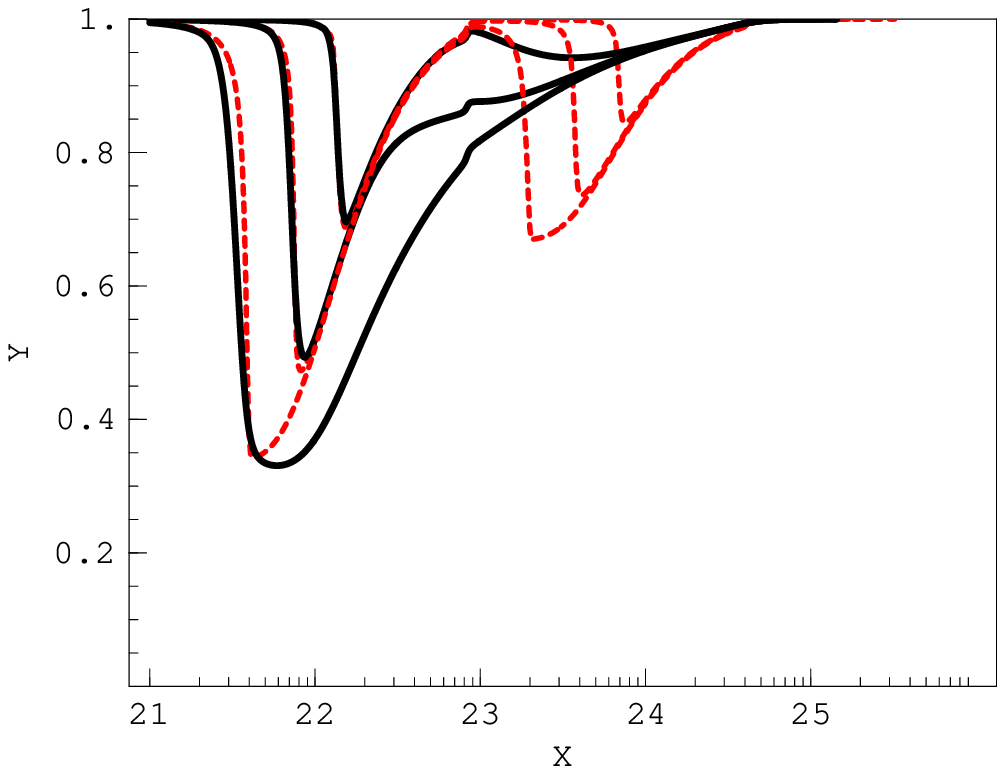,width=0.22\textwidth,angle=0}
 \hfill
 \psfig{file=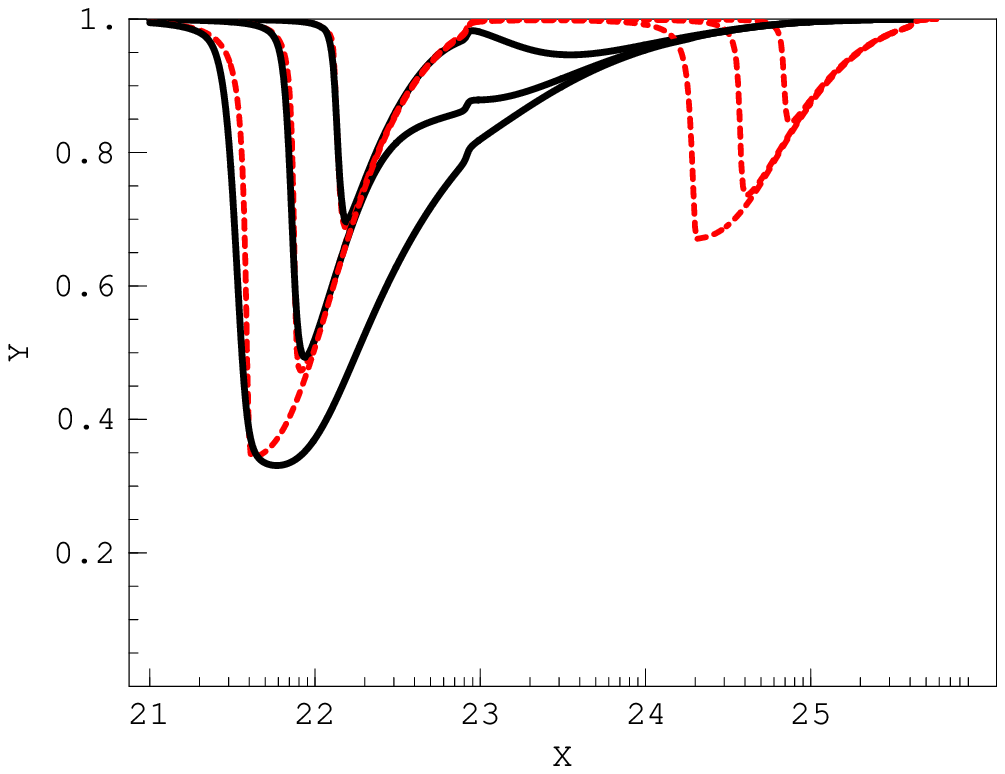,width=0.22\textwidth,angle=0}
 \caption{\footnotesize{Average transmission probability $\bar{P}_\mathrm{T}$ as a function of the 
\UHEnu energy as detected on Earth, $K_0$, for a source located at redshifts $z_\mathrm{s}=5$, 
$10$, $20$ (from top to bottom in each plot). The two left plots assume a normal hierarchy for the 
$m_\nu$, with values ($5\ 10^{-2}$ eV, $9\ 10^{-3}$ eV, $10^{-3}$ eV) and ($5\ 10^{-2}$ eV, $9\ 
10^{-3}$ eV, $10^{-4}$ eV) respectively. The right plots assume an inverted hierarchy with masses 
($5\ 10^{-2}$ eV, $5\ 10^{-2}$ eV, $10^{-3}$ eV) and ($5\ 10^{-2}$ eV, $5\ 10^{-2}$ eV, $10^{-4}$ 
eV) respectively. The continued, black curves correspond to the full damping as from 
eq.~(\ref{eq:gamma}) and (\ref{eq:sigma}), while the dotted (red) curves are for the approximation 
of relic neutrinos at rest.}} 
 \end{figure}
\section{Absorption lines due to relic neutrino clustering}
\label{sec:clustering}
\begin{figure}
 \begin{center}
 \psfrag{23}[c]{\tiny \phantom{n} \raisebox{0cm}{$10^{23}$}}
 \psfrag{22}[c]{\tiny \phantom{n} \raisebox{0cm}{$10^{22}$}}
 \psfrag{21}[c]{\tiny \phantom{n} \raisebox{0cm}{$10^{21}$}}
 \psfrag{20}[c]{\tiny \phantom{n} \raisebox{0cm}{$10^{20}$}}
 \psfrag{1.}[c]{\tiny \phantom{} \raisebox{0.1cm}{$1$}}
 \psfrag{0.8}[c]{\tiny \phantom{} \raisebox{0.1cm}{$0.8$}}
 \psfrag{0.6}[c]{\tiny \phantom{} \raisebox{0.1cm}{$0.6$}}
 \psfrag{0.4}[c]{\tiny \phantom{} \raisebox{0.1cm}{$0.4$}}
 \psfrag{0.2}[c]{\tiny \phantom{} \raisebox{0.1cm}{$0.2$}}
 \psfrag{0.}[c]{\tiny \phantom{} \raisebox{0.1cm}{$0$}}
 \psfrag{X}[c]{\tiny \raisebox{-0.5cm}{$K_0 [\mathrm{eV}]$}}
 \psfrag{Y}[c]{\tiny {$P_\mathrm{T}$}}
 \psfig{figure=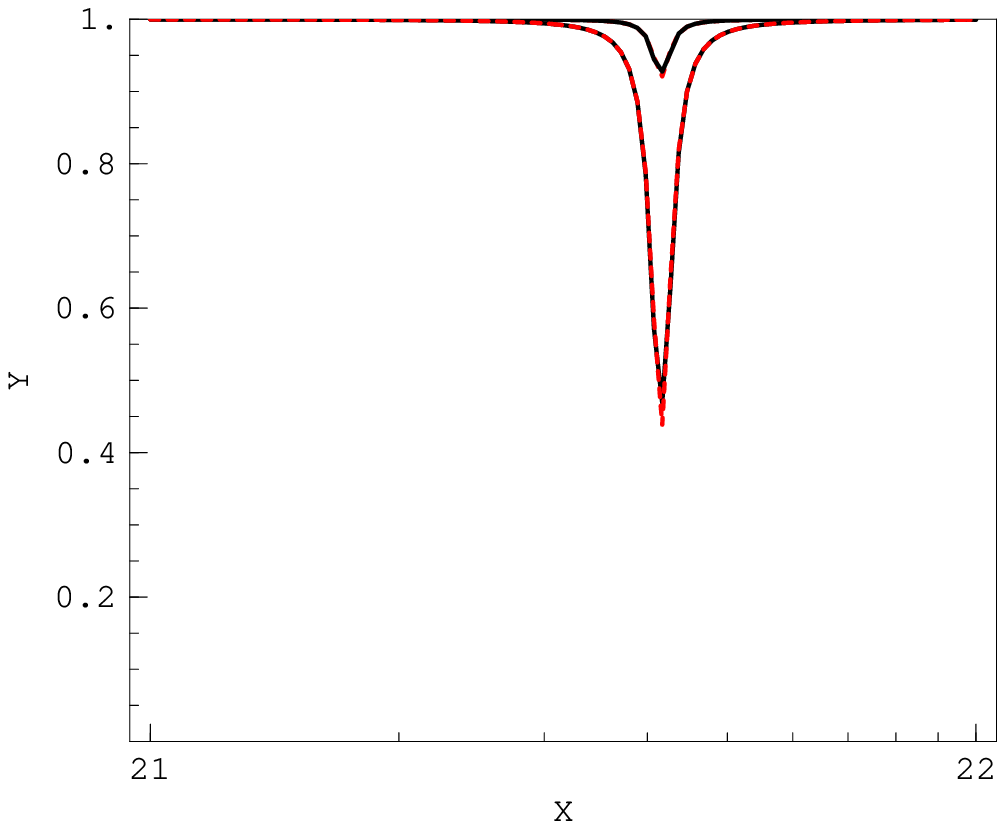,width=0.22\textwidth,angle=0}
 \hspace{2cm}
 \psfig{figure=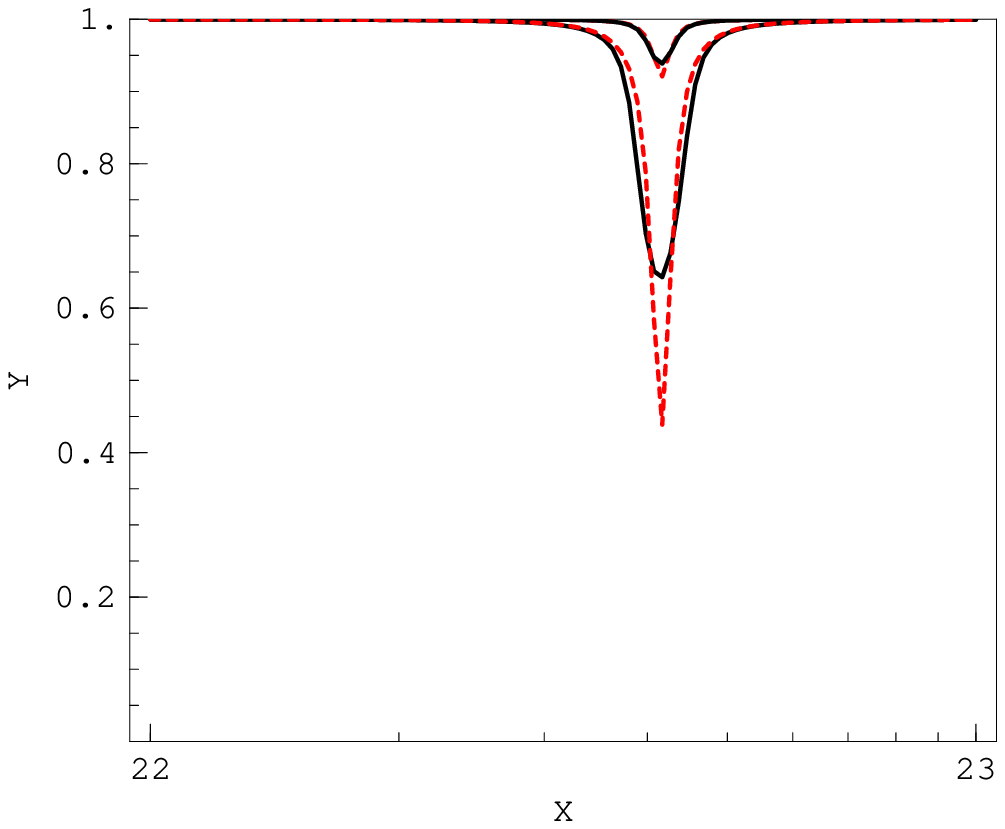,width=0.22\textwidth,angle=0}
 \hfill
 \caption{\footnotesize{Transmission probability for a cluster of extension 1 Mpc, made of 
neutrinos of mass 1 eV (left) and $10^{-1}$ eV (right), with a constant neutrino density 
$n^\mathrm{cl}_{\nu} = 10^3\ n_{0\nu}$ and $n^\mathrm{cl}_{\nu} = 10^4\ n_{0\nu}$ (from top to 
bottom in each plot). The colour code is the same as in fig.~1.  }}
\label{fig:cluster}
 \end{center}
 \end{figure}

According to recent calculations \cite{cluster}, depending on their mass and on the properties of 
the dark matter distribution, relic neutrinos cluster on scales ranging from $0.01$ to $1$ Mpc 
with overdensity factors $N_{cl}$ of the order of $10$ - $10^4$ with respect to the present 
background density ($n_{\nu 0} \approx 56 \mathrm{cm}^{-3}$ per species). To compute the \UHEnu 
absorption by clustered neutrinos we replaced $f_\nu(P)$ in eq.~\ref{eq:gammaUR} by a modified  
Fermi-Dirac distribution,
\begin{equation}
\label{eq:FDcluster}
f^{cl}_\nu(P) = \frac{1}{2}\frac{e^{-\Phi/T_\nu}+1}{e^{(P-\Phi)/T_\nu}+1},
\end{equation}
which fits reasonably well the profiles obtained in numerical calculations using Vlasov equation 
\cite{cluster}. 
For a given overdensity factor $N_{cl}$ we solve $\int_0^\infty \frac{dP}{2\pi^2} P^2 
f^{cl}_\nu(P) = N_{cl} n_{\nu 0}$ for $\Phi$. We computed the \UHEnu transmission probability 
across a cluster located between the \UHEnu source and the observer and found that the effect of 
the thermal motion of relic neutrinos is in general small or negligible. This is essentially 
because clustering is efficient only for neutrinos with mass $m_\nu \gtrsim 0.1$ eV, and because 
it is achieved only at small redshifts, when the \CnuB temperature is still very small. As shown 
in fig.~\ref{fig:cluster}, we have to saturate the bounds on the parameters to obtain a 
significant effect. For a maximal overdensity factor $N_{cl} = 10^4$, thermal effects reduce the 
maximum absorption probability across the cluster from $\approx 55 \%$ to $\approx 35 \%$.  

\section*{Acknowledgments}
We would like to acknowledge support by CONACyT under grants 34868-E
and 46999-F and by DGAPA-UNAM under grants PAPIIT IN116503, IN119405,
and IN112105.


\section*{References}
\begin{thebibliography}{15}
\bibitem{UHEnuCnuB} 
Weiler T J 1982 
{\it Phys.\ Rev.\ Lett.}\  {\bf 49} 234
\nonum Roulet E 1993 
{\it Phys.\ Rev.}\ D {\bf 47} 5247 
\bibitem{mass} Pas H and Weiler T J 2001
  {\it Phys.\ Rev.}\ D {\bf 63} 113015
\nonum Fargion D, De Sanctis Lucentini P G, Grossi M, De Santis M and Mele B 2002
  {\it Mem.\ Soc.\ Ast.\ It.}\  {\bf 73} 848  
\bibitem{ringwald}
Eberle B, Ringwald A, Song L, and Weiler T J 2004 
{\it Phys.\ Rev.}\ D {\bf 70} 023007
\bibitem{quigg} Barenboim G, Mena Requejo O and Quigg C 2005 
{\it Phys.\ Rev.}\ D {\bf 71} 083002
\bibitem{ourpaper} D'Olivo J C, Nellen L, Sahu S and Van~Elewyck V 2005 {\it Preprint} 
astro-ph/0507333  
\bibitem{dolivo} D'Olivo J F and Nieves J F 1995
{\it  Phys.\ Rev.}\ D {\bf 52} 2987
\bibitem{bell} Beacom J F and Bell N F 2002 
{\it  Phys.\ Rev.}\ D {\bf 65} 113009
\bibitem{cluster} Ringwald A and Wong Y Y Y 2004 
{\it JCAP} {\bf 0412} 005
\end{thebibliography}
\end{document}